\documentstyle[aps,multicol,psfig]{revtex}  

\begin{document}
\draft

\title{The magnetic response of disordered 
metallic rings: the large contribution of the far levels}

\author{M. Schechter$^1$, Y. Oreg$^2$, Y. Imry$^2$, and Y. Levinson$^2$}

\address{$^1$ The Racah Institute of Physics, Hebrew University, Jerusalem 
91904, Israel 
\newline
	$^2$Department of Condensed Matter Physics, The 
Weizmann Institute of Science, Rehovot 76100, Israel }

\maketitle


\begin{abstract}


We calculate the orbital linear magnetic response of disordered
metallic rings to an Aharonov-Bohm flux using the BCS model for
attractive electron-electron interaction. The contribution of all
levels including those up to a high energy cutoff results in a much
larger value than previously obtained using the local interaction
model. The possible relevance of our results to the resolution of the
discrepancy between the experimental and theoretical values for the
ensemble-averaged persistent currents in these systems is discussed.

\end{abstract}

\vspace{0.5cm}

\begin{multicols}{2}

One of the remarkable phenomena of mesoscopic physics is the existence
of equilibrium persistent currents in small normal metal rings, in the
regime where the elastic mean free path $l$ is much smaller than the
ring's circumference $L$.  This was both predicted theoretically
\cite{BIL83} and observed experimentally
\cite{LDDB90,CWB+91,MCB93,RRBM95,JMKW01,DBR+01}.  
While the phenomenon and the
periodicity in the flux are well understood by now, the magnitude of
the ensemble-averaged persistent current found experimentally is much
larger than that obtained theoretically using the model of
noninteracting electrons\cite{OR91,AGI91}. An attempt\cite{AE90,Sch91} to
account for this discrepancy by the inclusion of electron-electron
(``e-e'') interactions\cite{AL74} indeed increased the theoretical 
value, but still came short by a factor of about $5$\cite{AE90b}. 

Along with the large value of the ensemble-averaged 
persistent current observed, its sign poses significant questions as well. 
While in the first experiment\cite{LDDB90} the sign of the ensemble-averaged 
persistent current was tentatively identified as diamagnetic, it was 
not determined definitively 
(see Ref.~\onlinecite{AE90b}).  Later experiments reported a
predominantly diamagnetic ensemble-averaged persistent
current\cite{RRBM95,JMKW01,DBR+01}.  
Theoretically, the paramagnetic/diamagnetic
sign follows for repulsive/attractive effective e-e interactions (the
interplay between the renormalized\cite{MA62} repulsive e-e
interaction and the phonon mediated attractive interaction determines 
the sign of the effective interaction\cite{AE90b}).  
Thus, the diamagnetic sign of the persistent currents observed in 
Gold\cite{JMKW01}, Copper\cite{LDDB90}, and Silver\cite{DBR+01} 
suggests that these materials are very weak 
superconductors\cite{MM00}.

The model used to treat both signs of the interaction was that of
local e-e interaction. The discrepancy between the theoretical values
and the experimental results exists for repulsive\cite{AE90}, as well
as attractive interactions \cite{AE90b}. For the repulsive case, the
higher order renormalization which reduces the first order result is
responsible for this discrepancy. For the attractive interaction case,
the higher order renormalization does not reduce, and even somewhat
increases the first order result. However, the restriction on the bare
interaction constant, due to the experimental fact that Gold, Copper
and Silver are not superconducting in observed temperatures, limits
the value of the theoretical result in this case. For the magnetic
response (the derivative of the persistent current at zero flux),
there exist a further logarithm in the numerator\cite{AE90},
irrespective of the above renormalization of the effective
interaction. Here we show that the use of the BCS Hamiltonian to model
attractive e-e interaction gives rise to a much larger cutoff of this
logarithm, and hence to a significantly larger value for the magnetic
response to flux. The enhancement of the magnetic response is a result
of the contribution of high energy levels (to be called ``far
levels'') originating from pairing correlations which persist up to
the energy of the Debye frequency $\omega_{\rm D}$. Thus, correlations
on an energy scale much larger than the Thouless energy affect the
magnetic response.

The contribution of the far levels was investigated in connection to
superconductivity in small grains \cite{SILD01}. There, the
correlations of levels much further than the superconducting gap from
the Fermi energy $E_{\rm F}$, up to $\omega_{\rm D}$ prove to be
significant.  Using an exact solution of the reduced BCS Hamiltonian
\cite{Ric63,RS64}, it was shown that this contribution results in a
much larger condensation energy than that given by the BCS value, in a
wide parameter regime in which superconducting correlations are well
developed.  The far levels also affect single particle properties of a
superconducting grain\cite{SDIL02}, such as the Matveev-Larkin
parameter\cite{ML97}.

The difference between the local interaction model and the BCS
Hamiltonian lies in the $q$ dependence of the interaction in each,
where $q$ is the sum of the incoming (outgoing) momenta of the pair 
scattering. While the
local interaction model assumes that the bare interaction does not
depend on any momenta, and specifically on $q$, the BCS Hamiltonian
assumes a sharp cutoff at small total momentum, and considers the
interaction to be a delta function at $q=0$.  While for physical
properties related to the superconducting phase of the system, near and 
below $T_{\rm c}$ (such as
$T_{\rm c}$ itself)  the two models give similar results, we show that
for the magnetic response in the perturbative regime (equivalent to $T
\gg T_{\rm c}$) the results in the two models differ substantially. We
discuss the relation between the two models, and the effect of
relaxing the BCS assumption by taking the interaction to have finite
width in $q$.

Recently it was suggested that the largeness of the observed current
is due to AC noise \cite{Moh99,KA00} or interactions of the electrons with
two level systems \cite{Schw00}, and a relation to dephasing was suggested 
(one should also keep in mind that the latter interactions may lead to an 
additional attractive interaction). 
We, however, consider $T=0$, and do not include
dephasing (we consider the dephasing length $L_{\phi}$ to be larger
than all relevant lengths).  A finite $L_{\phi}$ will result in a
suppression of the persistent current, exponential in $L/L_{\phi}$.

We consider a quasi one dimensional
disordered ring penetrated by a constant Aharonov-Bohm (AB) flux $\Phi$ in
its middle.  We calculate the 
magnetic response, denoted by $F_2$, to first order in the e-e 
interaction, for attractive interaction using the reduced BCS Hamiltonian.

The result we obtain with logarithmic accuracy is 

\begin{equation}
\left\langle F_2 \right\rangle \equiv 
 \left\langle \left. \frac{dI}{d \Phi} \right|_{\Phi=0} \right\rangle = 
\frac{8 \pi \lambda E_{\rm Th}}{\Phi_0^2} 
\ln{\frac{E_{\rm co}}{d}} \; ,
\label{didphiresult}
\end{equation}  
where $\lambda$ is the dimensionless pairing parameter [see
Eq.~(\ref{HBCS})], $d$ is the level spacing, 
$E_{\rm Th} = l^2/(s \tau L^2)$ is the Thouless
energy, which is the inverse of the time to circulate the ring
($s=1,2,3$ is the effective dimension of the ring for diffusive
motion), $\Phi_0 = hc/(2e)$ and $\left\langle \cdot \cdot \cdot
\right\rangle$ denotes ensemble averaging. 
The upper logarithmic
cutoff $E_{\rm co} \equiv \min{[\omega_{\rm D},1/\tau]}$ represents
the fact that energies (measured from the $E_{\rm F}$) 
further than $E_{\rm Th}$, up to $E_{\rm co}$ (far levels) contribute 
to the magnetic response 
($\tau$ is the elastic mean free time).  Their
contribution enhances $F_2$ by about an order of magnitude due to the
much larger logarithmic cutoff as compared to $E_{\rm Th}$ of the
known result \cite{AE90} [for $1/(\tau E_{\rm Th}) \approx 10^4$
the ratio of the logarithms is roughly $10$].

In order to obtain Eq.~(\ref{didphiresult}) we consider the general 
Hamiltonian 

\begin{equation}
H = \sum_{\alpha} \int dr \, \psi^{\dagger}_{\alpha} (r) 
\left[\frac{1}{2m}(\vec{P} - \frac{e}{c}\vec{A})^2 + 
U(r)\right] \psi^{}_{\alpha} (r) + H_{\rm ee} \, ,
\label{generalH}
\end{equation} 
where $U(r)$ is the external potential which includes the disorder,
and $H_{\rm ee}$ represents the e-e interaction. 
In the London gauge $\vec{A} = \Phi/(2 \pi \rho) \hat{\phi} \;$
where $\rho$ is the distance from the origin and 
$\hat{\phi}$ is in the clockwise direction of the ring.  The free
energy of the system and the persistent current are flux dependent, and 
related by $I(\Phi)=-dF/d\Phi$. By time reversal symmetry 
$I(0) = 0$, and for small flux 

\begin{equation}
F(\Phi) = F_0 - \frac{1}{2} F_2 \Phi^2 + ... \; ; \qquad 
\left. \frac{dI}{d \Phi} \right|_{\Phi=0} = F_2 \; . 
\label{EPhi}
\end{equation} 

We now turn to the calculation of $F_2$ to first order in the
interaction. Perturbative analysis of the reduced BCS attractive interaction 
[Eq.~(\ref{HBCS}) below] is valid, at $T=0$, for 
$\lambda < 1/\ln(\omega_{\rm D}/d)$\cite{SILD01} 
and at finite temperature ($T > d$) for $T \gg T_{\rm c}$. 
We take as the unperturbed Hamiltonian

\begin{equation}
H_0 = \sum_{\alpha} \int dr \, \psi^{\dagger}_{\alpha} (r) 
\left[\frac{P^2}{2m} + U(r) \right] \psi^{}_{\alpha} (r)
\label{Hzero}
\end{equation}
and the magnetic field together with the e-e interaction as perturbation 

\begin{eqnarray}
H_I =  H_{\rm ee} + & &  \\ 
\sum_{\alpha} \int dr \, \psi^{\dagger}_{\alpha} (r) & 
\left[ - \displaystyle\frac{e}{2 m c}(\vec{P} \cdot \vec{A} + \vec{A} \cdot \vec{P}) + 
\frac{e^2}{2 m c^2} A^2 \right] \psi^{}_{\alpha} (r) &  \, .  \nonumber 
\end{eqnarray} 
We calculate perturbatively the energy to third order in $H_{\rm I}$,
and consider only the terms which are second order in the flux and
first order in the interaction.

We denote by $| i \rangle$ the eigenstates of the noninteracting
electrons in the disordered ring without magnetic field.  In this
basis, within the reduced BCS model, we obtain

\begin{equation}
H^{\rm BCS}_{\rm I} = - \sum_{i j \alpha} \frac{e \Phi}{m c L} P_{ij} 
c^\dagger_{i \alpha} c^{}_{j \alpha}  + 
\lambda d \sum_{ij} \, \hspace{-0.1cm}' c^\dagger_{i \uparrow} 
c^\dagger_{i \downarrow} c^{}_{j \downarrow} c^{}_{j \uparrow} \, .
\label{HBCS}
\end{equation} 
Here $c_i$ destroys an electron in the state $| i \rangle$ with
wavefunction $\chi_i (r)$ and $P_{ij}=\langle i |P_{\|}| j \rangle$ is
the matrix element of the momentum parallel to the ring's direction. 
The second sum is restricted to levels within $\omega_{\rm D}$ 
of $E_{\rm F}$. 
Note that contrary to the usual convention, $\lambda$ is negative for
attractive interaction.  
We assume that the width of the ring is much smaller
than its radius \cite{fluxfield}. 
The $A^2$ term is then interaction independent. 
We choose the $\chi_i$'s to be real, and then $P_{ij}$ is pure
imaginary and $P_{ii}=0$.

Using third order perturbation theory \cite{LL91} we find that for
$T=0$ to first order in $\lambda$

\begin{equation}
F_2 = 8 \left( \frac{e}{mcL} \right)^2 \lambda d \hspace{-0.3cm} 
\sum_{ \vspace{-0.4cm}
\begin{array}{c} 
\vspace{-0.45cm}
\\ {\scriptstyle 
0<i<\omega_{\rm D} } 
\vspace{-0.1cm} 
\\ 
{\scriptstyle   -\omega_{\rm D}<j<0 } \end{array}} \hspace{-0.4cm}
\frac{|P_{ij}|^2}{\omega_{ij}^2} 
\, ,
\label{E2lam}
\end{equation} 
where $0<i<\omega_{\rm D}$ represents states whose energy is between
$E_{\rm F}$ and $E_{\rm F}+\omega_{\rm D}$, and
$\omega_{ij}=\epsilon_i - \epsilon_j$.  Performing disorder averaging
over $|P_{ij}|^2$ we obtain 
\begin{equation}
\left\langle |P_{ij}|^2 \right\rangle = 
\frac{p_{\rm F}^2 d \tau}{\pi 
(1 + \omega_{ij}^2 \tau^2) s} \, ,
\label{pijaverage}
\end{equation}
which is roughly constant for $\omega_{ij} < 1/\tau$ and zero for
$\omega_{ij} > 1/\tau$.  In Eq.~(\ref{E2lam}), for energies smaller
than $\omega_{\rm D}$ and $1/\tau$ there is a double sum and a second
power of energy in the denominator. The result is therefore a
logarithm with an upper cutoff which is given by $\min{[\omega_{\rm
D}, 1/\tau]}$.  We evaluate the sum in Eq.~(\ref{E2lam}) taking for
simplicity the noninteracting spectrum to be equally spaced, and
obtain Eq.~(\ref{didphiresult}).

This equation differs from the result of Ambegaokar and Eckern 
(AE)\cite{AE90} by the upper cutoff 
of the logarithm, being  $\min{[\omega_{\rm D},1/\tau]}$ in comparison to 
$E_{\rm Th}$ in Ref.~\onlinecite{AE90}. We now explain the origin of 
this difference in some detail. 
AE consider the same Hamiltonian as in our 
Eq.~(\ref{generalH}), only with local e-e interaction 
$V(r_1-r_2) = \tilde\lambda N(0)^{-1} \delta(r_1-r_2)$. 
In terms of the noninteracting eigenstates $H_{\rm ee}$ is then given by

\begin{equation}
H^{\rm local}_{\rm ee} = \frac{1}{2} 
\sum_{ijkl,\sigma} V_{ijkl}  c^\dagger_{i\sigma} 
c^\dagger_{j\bar{\sigma}} c^{}_{k\bar{\sigma}} c^{}_{l\sigma} \; ,
\label{Heelocal}
\end{equation} 
where $V_{ijkl} = \tilde\lambda N(0)^{-1}
\int \chi_i(r) \chi_j(r) \chi_k(r) \chi_l(r) dr$ 
and $\sigma \neq \bar{\sigma}$ due to the interaction being
local. The dependence of the energy on flux, to first order in the 
interaction, can be written as 

\begin{equation}
\Delta F = \sum_{i,j<0} V_{ijji}(\Phi) = 
\sum_{i \neq j<0} V_{ijji}(\Phi) + \sum_{i<0} V_{iiii}(\Phi) \; .
\label{ijji}
\end{equation} 
We now show that the second, diagonal part contribution to the magnetic 
response corresponds to the BCS result as given in Eq.~(\ref{didphiresult}). 
$\Delta F$ can be expanded to second order in the flux to obtain 
$F_2$ for the local interaction model by following the same procedure 
done above for the reduced BCS Hamiltonian, with $H_{\rm ee}$ given by 
Eq.~(\ref{Heelocal}). The result is 

\begin{equation}
\left\langle F_2 \right\rangle = 4 \left( \frac{e}{m c L} \right)^2 
\left\langle \sum_{j<0, i>0} 
\frac{|P_{ij}|^2}{\omega_{ij}^2} \sum_{l<0} (V_{jllj} - V_{illi}) 
\right\rangle \; .
\label{localres}
\end{equation}
The diagonal contribution to this result is given by the $l=i$ terms, and 
since to leading order $V_{iiii} - V_{ijji} =2 \tilde\lambda d$\cite{Mir00} 
we obtain for 
the diagonal contribution in the local interaction model the same expression 
as Eq.~(\ref{E2lam}), without the explicit cutoff at $\omega_{\rm D}$. 
Due to Eq.~(\ref{pijaverage}), the diagonal contribution gives 
Eq.~(\ref{didphiresult}) with  $E_{\rm co} = 1/\tau$. 
Therefore, the BCS approximation is equivalent 
to assuming that the flux dependence of the offdiagonal matrix elements is 
small, and can be neglected. In the case of the local interaction model 
this is not the case, and the contribution of the offdiagonal elements to 
the magnetic response is significant, and opposite in sign to that of the 
diagonal element. This results in a partial cancellation, and therefore 
a reduction of the high logarithmic cutoff in Eq.~(\ref{didphiresult}).

Another way to understand the relation between the two models is to consider 
the $q$ dependence of the interaction in both (where $q$ is the sum of the 
incoming momenta).  
AE derive their result as a sum over all $q$'s (see Eqs.~[(12)-(14)] of 
Ref.~\onlinecite{AE90}), 
and the result within the BCS Hamiltonian [Eq.~(\ref{didphiresult})] 
corresponds to their $q=0$ term. For the local interaction 
model, the contribution to the magnetic response 
of the high energies, above $E_{\rm Th}$ exactly 
cancels between equal magnitude and opposite signs of the $q=0$ term and 
the sum of all $q \neq 0$ terms. \emph{This makes the $q$ independent 
assumption for the bare interaction crucial}. 
The existence of excess interaction at 
small total momentum $q$ would thus significantly affect the result.
The BCS interaction assumes just that, the existence of 
excess interaction at $q=0$. 

A physical justification for taking a $q$ dependent interaction 
can be obtained, 
for the attractive phonon mediated interaction, from the 
usual restriction that all (incoming and outgoing) scattered states are 
within $\omega_{\rm D}$ of $E_{\rm F}$ (see e.g. Ref.~\onlinecite{AGD63}). 
This restriction implies a significant $q$ dependence on the scale of 
$q_c=\omega_{\rm D}/v_{\rm F}$. One can therefore take a $q$ dependent 
attractive interaction in the form of a step function with width $q_c$. 
For $q_c < 2 \pi/L$ one finds that only the $q=0$ term survives, and the BCS 
result for the magnetic response is recovered. 

For metals such as Gold, Copper and Silver, this condition requires 
the circumference of the ring to be smaller than $0.2-0.4 \mu m$, which 
is an order of magnitude smaller than the relevant experimental 
lengths\cite{LDDB90,RRBM95,DBR+01}. 
The effects of relaxing this condition, i.e. having $q_c \gtrsim 2 \pi/L$, 
as well as the effect of the $q$ dependence on the persistent current itself 
will be considered elsewhere\cite{SOIL02b}. 

The central result of this Letter is the large interaction correction
to the derivative of the persistent current at zero flux, within the 
BCS model, as is given in Eq.~(\ref{didphiresult}). 
Physically, this is due to the large contribution of the far levels, up 
to $\min{[\omega_{\rm D},1/\tau]}$ from the Fermi level.  

Our calculations were done for $T=0$. 
At finite temperature $T < \min{[\omega_{\rm D},1/\tau]}$ 
(but neglecting dephasing, assuming $L_\phi \gg L$) the magnetic response 
would be given by Eq.~(\ref{didphiresult}) with $d$ replaced by $T$. 
Therefore, on top of the large magnetic response, we predict a weak, 
logarithmic, temperature dependence of the magnetic response up to $T 
\approx \min{[\omega_{\rm D},1/\tau]}$.  The weak temperature 
dependence is due to the addition of the many small contributions, all
having the same sign, of the levels up to the large energy cutoff (see
also Refs.\onlinecite{SILD01,LFH+00}).

The local interaction model and the reduced BCS Hamiltonian are two models 
used to describe the effective e-e interaction resulting from the retarded 
electron-phonon interaction. Though different, these two models give similar 
results for many properties of the superconductor near and below $T_{\rm c}$. 
The robustness of the physics near and below the transition temperature 
makes the differences between the two models irrelevant. However, the 
magnetic response in the perturbative regime is a more subtle property, 
that distinguishes between the two models. 

Our result for the reduced BCS Hamiltonian also follows by expanding 
Richardson's exact solution\cite{Ric63}. This has the analytical merit 
of being a first order expansion of an exact solution, as well as an 
additional viewpoint on the contribution of the far levels, and will 
be given elsewhere\cite{SILO02c}.

We would like to thank O. Agam and I. Aleiner for discussing with us 
the special role of the $q=0$ term in the calculation of the 
persistent current within the local interaction model. 
We benefited from valuable discussions with 
N. Argaman, V. Kravtsov, D. Orgad, A. Punnoose, Y. Aharonov, B. Altshuler, 
R. Berkovits, H. Bouchiat, P. Brouwer, J. von Delft, 
A. Finkel'stein, Y. Gefen, D. E. Khmelnitskii, M. Khodas, 
F. von Oppen, Z. Ovadyahu, R. Smith, U. Smilansky, R. A. Webb, 
and H. A. Weidenmueller. 
M.S. is thankful for the support by the Lady Davis fund. 
This work was supported by a Center of Excellence of the Israel
Science Foundation (ISF), Jerusalem and by the German Federal Ministry of
Education and Research (BMBF) within the Framework of the
German-Israeli Project Cooperation (DIP), by the German-Israeli
Foundation(GIF), and by ISF grant number 160-01/1.

\end{multicols}


\begin{thebibliography}{10}
\bibitem{fluxfield}
Under this assumption of thin ring the response to constant magnetic field can
  be found along the same lines as done here for AB flux.
\end{thebibliography}
\end{document}